\documentclass[11pt,reqno,a4paper]{amsart}

\usepackage[UKenglish]{babel}
\usepackage[table]{xcolor}
\usepackage{acro}[=v2]
\usepackage{booktabs}
\usepackage{csquotes}
\usepackage{graphicx} 
\usepackage{enumerate}
\usepackage{mathptmx}
\usepackage{bbm}
\usepackage{multirow}
\usepackage{wasysym}
\usepackage{float}
\usepackage{cancel}
\usepackage[shortlabels]{enumitem}
\usepackage{adjustbox}
\usepackage{multicol}
\usepackage{booktabs}
\usepackage{url}
\usepackage[outdir=./]{epstopdf} 
\usepackage{subcaption}
\usepackage[foot]{amsaddr}

\usepackage[shortlabels]{enumitem}

\usepackage{hyperref}
\hypersetup{
    colorlinks=true, 
    linktoc=all,     
    linkcolor=blue,  
}

\usepackage[left=4cm,right=4cm,top=3cm,bottom=3cm,headsep=0.6 cm,footnotesep=0.6cm]{geometry}
\usepackage[size=tiny, textwidth=2cm]{todonotes}

\usepackage{amssymb,amsmath,amsfonts,mathtools,color}

\usepackage{mathrsfs}
\usepackage{bm} 

\usepackage{comment} 
\usepackage[title,titletoc,header]{appendix}

\usepackage[linesnumbered, ruled,vlined]{algorithm2e}
\SetKwInput{KwInput}{Input}                

\graphicspath{ {./figures/} }

\numberwithin{equation}{section}

\newtheorem{Theorem}{Theorem}[section]

\newtheorem{Assumption}{Assumption \!\!}

\theoremstyle{definition}
\newtheorem{Definition}{Definition}[section]

\theoremstyle{remark}

\def\cF{\mathcal{F}}

\def\cT{\mathcal{T}}

\def\d{{\mathrm{d}}}

\def\sF{{\mathbb{F}}}

\def\sP{\mathbb{P}}
\def\sQ{{\mathbb{Q}}}

\newcommand{\lc}
{\mathrel{\raise2pt\hbox{${\mathop<\limits_{\raise1pt\hbox
{\mbox{$\sim$}}}}$}}}

\newcommand{\gc}
{\mathrel{\raise2pt\hbox{${\mathop>\limits_{\raise1pt\hbox{\mbox{$\sim$}}}}$}}}

\newcommand{\ec}
{\mathrel{\raise2pt\hbox{${\mathop=\limits_{\raise1pt\hbox{\mbox{$\sim$}}}}$}}}

\def\bb{\begin{equation}} \def\ee{\end{equation}}
\def\bbn{\begin{equation*}} \def\een{\end{equation*}}

\def\beqn{\begin{eqnarray}}  \def\eqn{\end{eqnarray}}

\def\beqnx{\begin{eqnarray*}} \def\eqnx{\end{eqnarray*}}

\def\bn{\begin{enumerate}} \def\en{\end{enumerate}}

\def\bd{\begin{description}} \def\ed{\end{description}}



\title{Pricing and hedging of decentralised lending contracts  }

\author{Lukasz Szpruch$^{1,2,3}$}
\author{Marc Sabat\'e Vidales$^{1,2}$}
\author{Tanut Treetanthiploet$^{4}$}
\author{Yufei Zhang$^5$}
\address{$^1$Simtopia.ai}
\address{$^2$School of Mathematics, University of Edinburgh}
\address{$^3$The Alan Turing Institute}
\address{$^4$Infinitas by Krungthai, Bangkok}
\address{$^5$Imperial College London}
\thanks{\textbf{Acknowledgements:} We would like to thank Samuel N. Cohen (University of Oxford), and David Siska (University of Edinburgh \& Vega Protocol) for their feedback on this work. }

\date{}

\begin{document}
\maketitle

\begin{abstract}
We study the loan contracts offered by decentralised loan protocols (DLPs) through the lens of financial derivatives. DLPs, which effectively are clearinghouses, facilitate transactions between option buyers (i.e. borrowers) and option sellers (i.e. lenders). The loan-to-value at which the contract is initiated determines the option premium borrowers pay for entering the contract, and this can be deduced from the non-arbitrage pricing theory. We show that when there are no market frictions, and there is no spread between lending and borrowing rates, it is optimal to never enter the lending contract.  

Next, by accounting for the spread between rates and transactional costs, we develop a deep neural network-based algorithm for learning trading strategies on the external markets that allow us to replicate the payoff of the lending contracts that are not necessarily optimally exercised. This allows hedge the risk lenders carry by issuing options sold to the borrowers, which can complement (or even replace) the liquidations mechanism used to protect lenders' capital. Our approach can also be used to exploit (statistical) arbitrage opportunities that may arise when DLP allow users to enter lending contracts with loan-to-value, which is not appropriately calibrated to market conditions or/and when different markets price risk differently. We present thorough simulation experiments using historical data and simulations to validate our approach.   
\end{abstract}

\section{Introduction}
  
  Decentralised lending protocols (DLPs) resemble a collateralised debt market (CDM) by pooling assets from lenders to enable over-collateralised loans to borrowers without having to rely on a central trusted entity \cite{harvey2021defi,schar2021decentralized,xu2023sok,john2022smart}. Protocol governance needs to monitor current market conditions to decide maximum loan-to-value ratios at loan origination for each pair of assets, which dictate how much of the debt asset can be borrowed at time zero, given posted collateral. One way to decide maximum loan-to-value ratios is to use statistical approaches involving coherent risk measures, which have been developed in \cite{cohen2023paradox} and which aim to control the probability of lenders losing assets and/or the amounts lost in the case of default happens. 
  
In this work, we study the loan contracts offered by DLPs through the lens of financial options. DLP - a clearinghouse - facilitates transactions between option buyers (i.e. borrowers) and option sellers (i.e. lenders). For example, protocols such as Aave \cite{aave}, Compound \cite{compound}, or Morpho \cite{morpho} offer loan contracts that resemble stock loans in TradFi, which essentially are American perpetual barrier option with a barrier above the strike as has been demonstrated in \cite{szpruch2024leveraged}. The loan-to-value at which the contract is initiated determines the option premium borrowers pay for entering the contract and it can be deduced from the non-linear non-arbitrage pricing theory \cite{el1997backward}, where non-linearity in pricing arises due to different rates for lending and borrowing\footnote{A textbook non-arbitrage pricing theory makes a simplifying assumption that agent can lend and borrow at the same risk free interest rate. This streamlines the analysis and lead to linear pricing theory. In contrast, when lending and borrowing rates differ, it is not even clear what the correct discount rate should be, and pricing equations become nonlinear. See \cite{el1997backward} for gentle introduction to non-linear pricing theory }. There are a number of consequences of analysing lending contracts using non-arbitrage pricing theory:
  \begin{enumerate}
      \item \textbf{Non-arbitrage loan-to value}. Non-arbitrage consideration, and law of one price in particular, tell us that portfolios generating the same cash flows ought to have the same initial value. By building replicating portfolios on external spot and derivatives markets that mirror cash flows generated by loan contracts, one can exploit possible arbitrage opportunities that may arise when loan contract position is established with the initial loan-to-value not appropriately calibrated to market data and / or when participants in different markets value risks differently. 
      \item \textbf{Risk management of DLP}. On many DLPs, debt positions which become not sufficiently collateralised are auctioned off to liquidators at a discount. This design suffers from the paradox of adversarial liquidations \cite{cohen2023paradox}, and liquidations spirals arise during periods of high volatility / thin liquidity \cite{cohen2023paradox,qin2023mitigating}. Using insights developed in this work one can build hedging strategies on the external markets to hedge the risk lenders carry by issuing options sold to the borrowers, which can complement (or even replace) the liquidations mechanism. 
      \item \textbf{Mechanism design for lending protocols}. Careful design is required in order to provide the right incentives to participants and to maintain a stable balance of DLPs users under varied economic and market conditions and scenarios. Therefore it useful to have a generic framework that can encompass multiple design choices. By viewing lending contracts as financial derivatives and DLPs as corresponding clearing houses one can systematically evaluate multiple design decisions as this will translate into corresponding options payoffs and clearing rules.
  \end{enumerate}


\subsection{Literature review}

Lending contracts are similar to stock loans and have been studied in the literature through the lense of american options in  
\cite{xia2007stock,grasselli2013stock,mcwalter2022stock}. A novel feature of many DLPs, compared to stock loans, is that debt positions which are not sufficiently collateralised are auctioned off to liquidators at a discount. This means that the holder of the loan effectively holds an American perpetual barrier option with a barrier above the strike. Viewing lending contracts as perpetual options provides a basis for the design of DLPs, which involves the choice of loan-to-value ratios at loan origination, liquidation thresholds, and bonuses, which control the acceptable level of risk for the protocol and its users, and also collateralisation rules, which in turn control the exposure of the protocol to a particular asset class. We refer the reader to \cite{cohen2023paradox} for a systematic analysis of these design choices. There are also a number of empirical works that shed light on risk and reward trade-offs in DLPs \cite{qin2021empirical,perez2021liquidations,lehar2022systemic}. There is also a growing literature on studying market efficiency and equilibrium in the context of lending protocols, \cite{cohen2023economics, rivera2023equilibrium,chiu2022inherent,chaudhary2023interest,mueller2023defi,aramonte2022defi,chitra2023attacks,bertucci2024agents}.

\section{Description of lending protocol}

Consider a lending   mechanism between ETH and USDC.
Let $(P_t)_{t\geq 0}$ denote the price process of a risky asset, which we set to be ETH, with a dollar stablecoin being the num\'eraire.  We take that stablecoin to be USDC, so for all $t\geq 0$,  $1\text{ETH}=P_t\text{USDC}$.  This is a standard accounting convention that does not take into account market frictions. 

 Let $(r^{b,E}, r^{c,E})$, $(r^{b,D}, r^{c,D})$ be interest rates for borrowing and providing collateral for ETH and USDC respectively.  Furthermore, let $\theta^0\in[0,\text{maxLTV})$ be an initial loan-to-value.  Loan-to-value corresponds to the initial haircut on the collateral, which is less or equal to the maximal loan-to-value allowed by a protocol, meaning one borrows $\theta^0$ ETH of value for every unit of ETH deposited as collateral. 
 
 \subsection{Down-and-out American barrier option via Lending protocol}\label{sec down and out American barrier option}
 We first analyse lending contract from the borrower perspective.
 To open a long-ETH loan position, an agent 1) purchases 1 ETH on the market for $P_0$ of USDC, 2) deposits 1 ETH as collateral, 3) borrow $\theta \,P_0$ USDC against the collateral. We see that effectively only $P_0(1 - \theta^0)$ is required to establish the position. For capital efficiency the agent may use a flashswap~\footnote{This takes advantage of atomicity of the transactions executed on chain, but means the agent is exposed to slippage on a DEX of choice. When using a standard loan the ETH can be purchased on a centralised, potentially more liquid, exchange.} (or a flashloan and a swap) along the following steps: 
\begin{enumerate}
\item Begin with $P_0(1-\theta^0)$ of USDC.
\item Obtain 1 ETH using flashswap (need to deposit $P_0$ USDC within one block for this to materialise)\footnote{One can also get USDC via flashloan and swap it for ETH, but this involves additional gas fee, fee for trading and suffer from temporal market impact}.
\item Deposit 1 ETH as collateral and start earning interests according to $e^{r^{c,E}\,t}$. If there is no rehypothecation of collateral, $r^{c,E}=0$.
\item Borrow $\theta^0 \,P_0$ of USDC against the collateral and start paying interests according to $\theta^0 \,P_0 e^{r^{b,D}\, t}$
\item Put together $\theta^0 \,P_0$ and initial amount $P_0(1-\theta^0)$  of USDC to complete the flashswap. 
\end{enumerate}

The loan position has a maturity $T>0$.
At any time $t\in [0,T]$, the holder of the position may choose to pay back the loan $\theta^0 \,P_0 e^{r^{b,D}\, t}$ in exchange for the collateral with value $P_t\,e^{r^{c,E}\,t} $. Note that a rational agent will only do that if $\theta^0 \,P_0 e^{r^{b,D}\, t} \leq P_t\,e^{r^{c,E}\,t}$, otherwise it is better to walk away from the position.  Hence,   the agent in entitled to the  payoff 
\begin{equation}\label{eq call option}
  (P_t\,e^{r^{c,E}\,t} -\theta^0 \,P_0 e^{r^{b,D}\, t} )^{+}\,,
\end{equation}
 where $x^+ = \max\{0,x\}$.

 Many leading protocols have liquidation constraints.
 If the value of the asset falls too low, the position will be liquidated.
 Let $\theta \in (\theta^0,1]$ be the liquidation loan-to-value (LLTV),
 and let $\tau^B$ be the liquidation time  defined by 
\begin{equation}
\label{eq:liquidation_time}
   \tau^B \coloneqq \inf \left\{ t \in [0,T] \mid \theta P_t e^{r^{c,E} \, t} \leq \theta^0 \,P_0\ e^{r^{b,D} \,t}  \right\}\,.
\end{equation}
Since LLTV $\theta<1$ then for all $t<\tau^B$
\begin{equation*}
0< \theta P_t e^{r^{c,E} \, t} - \theta^0 \,P_0\ e^{r^{b,D}} < P_t e^{r^{c,E} \, t} -  \theta^0 \,P_0\ e^{r^{b,D}}\,.
\end{equation*}
The payoff accounting for liquidations\footnote{For simplicity, we assume that once the position is open for liquidation, it disappears from the agent balance sheet.} is given by 
\begin{equation}\label{eq payoff}
(P_t\,e^{r^{c,E}\,t} - \theta^0 \,P_0\,e^{r^{b,D}\, t} )\mathbbm{1}_{\{t<\tau^B\}}\,.
\end{equation}
The initial capital needed to enter the lending position, which we refer to as lending contract premium, is given by $P_0(1-\theta^0)$. From non-arbitrage consideration, $P_0(1-\theta^0)$ should be the exact amount needed to establish a self-financing replicating portfolio, that is, portfolio such that at every time $t$ its value matches the payoff \eqref{eq payoff}. Typically, the replicating portfolio consists of trading on the money market, where USDC yields interests, and on the spot market. However, one can also consider other portfolios that consist of statically and/or dynamically traded risky assets or their derivatives. 

  Note that the lending contract with payoff \eqref{eq payoff} is  equivalent to a down-and-out barrier option, where the position becomes worthless to its holder when the value of the collateral falls sufficiently low. Note that up until the liquidation, the pay-off of this option is linear in the underlying risky asset, making it closely related to perpetual futures, with the barrier being a manifestation of margin closeouts with the appropriately chosen maintenance margin rule. This is studied in \cite{szpruch2024leveraged}. On the other hand, when LLTV is close to one, the lending contract is similar to a perpetual call option with a time-dependent strike with underlying being given by  
  $P e^{r^{c,E}}$, which can be thought of as a risky asset with price $P$ paying dividend at the rate $r^{c,E}$ . In particular when LLTV is equal $1$ or equivalently when there are no liquidations, the lending contract is precisely a perpetual call with payoff given by \eqref{eq call option}.  

In this work, we aim to answer the following questions: 
\begin{enumerate}
    \item Are lending contracts mispriced? That is, given fixed $\theta$ and $\theta^0$, does $P_0(1-\theta^0)$ correspond to the non-arbitrage value of the contract with the payoff \eqref{eq payoff} under varying market conditions? 
    \item By accounting for market frictions what is the initial capital required to establish a portfolio replicating the payoff \eqref{eq payoff} and how one can efficiently learn the corresponding trading strategy using modern deep learning tools? 
\end{enumerate}

\subsection{Notation}
 We shall assume
that $(\Omega,\cF, \sP)$
 is a complete probability space, and 
 $\sF=(\mathcal F_t)_{t\ge 0}$
 is the filtration generated by the non-negative continuous price process $P$, 
augmented by $\sP$-null sets. By $\mathbb Q$ we denote any measure such that $(e^{-r^{c,D}} P_t e^{r^{c,E} t})_t$ is $\mathbb Q$-martingale. That is, under $\mathbb Q$ the discounted dividend-yielding process $P$ is a martingale.  Let $\mathcal T$ be the set of stopping times taking values in $[0,\infty)$.

\section{Borrowing and lending without rate spreads. Explicit formulae.} \label{no spread}

In this section, we derive the fair price of the lending contract \emph{from the borrower's perspective} in the case when interests for lending and borrowing are the same i.e.,  
   $r^{b,D}=r^{c,D}$ and 
  $r^{b,E}=r^{c,E}$, and there are no market frictions. 
  \begin{Assumption}\label{as no spread}
  We assume that there exists an external market where:
\begin{itemize}
\item The agent can borrow and lend any amount of USDC at the riskless rate $r^{c,D}$.
\item The agent can buy and sell any amount of risky asset with price $P$. 
\item The above transactions do not incur any transaction costs and the size of a trade does not impact the prices of the traded assets. 
\end{itemize}
\end{Assumption}
This case allows for the derivation of the analytical formula for the no-arbitrage price of the option with the payoff \eqref{eq payoff}, which we derive in the Appendix \ref{sec price formula without spread}. 
The challenging case of pricing and hedging under different rates for borrowing and lending and accounting for market frictions will solved using deep learning techniques in the next section. Due to computation burden, such computation typically will have to be done off-chain. 

First, we observe that interest paid on collateral, represented here by risky asset $P$, is equivalent to the asset paying a dividend with the rate $r^{c,E}$. Classical non-arbitrage pricing theory applied to risky asset yielding dividends, tells us  that absence of arbitrage implies the existence of martingale measure $\mathbb Q$ such that $(e^{-r^{c,D}} P_t e^{r^{c,E} t})_t$ is $\mathbb Q$-martingale.

\subsection*{Loan contracts are overpriced.}

 Let $\mathcal T$ be the set of stopping times taking values in $[0,\infty)$.
Let  $\tau \in \mathcal T$ be a stopping time 
at which  the holder chooses to pay back the loan. The non-arbitrage price of this contract is given by
\begin{equation}
\label{eq:nonarb_price_nospread}
\sup_{\tau \in \mathcal T} \mathbb E^{\mathbb Q} \left[ e^{-r^{c,D} \tau}(P_\tau \,e^{r^{c,E}\,\tau } -\theta^0 \,P_0 e^{r^{c,D}\, \tau } )\mathbbm{1}_{\{\tau<\tau^B\}} \right]\,, 
\end{equation}
 where 
\begin{equation}
\label{eq:liquidation_time}
   \tau^B \coloneqq \inf \left\{ t \in [0,\infty) \mid \theta P_t e^{r^{c,E} \, t} \leq \theta^0 \,P_0\ e^{r^{c,D} \,t}  \right\}\,.
\end{equation}
\begin{Theorem}\label{th non-arbitrge price}
Let Assumption \ref{as no spread} hold. Let $\mathbb Q$ be such that $(e^{-r^{c,D}} P_t e^{r^{c,E} t})_t$ is $\mathbb Q$-martingale. Then it is optimal to exercise the loan contract at time $\tau^*=0$ and consequently
\begin{equation}
\sup_{\tau \in \mathcal T} \mathbb E^{\mathbb Q} \left[ e^{-r^{c,D} \tau}(P_\tau \,e^{r^{c,E}\,\tau } -\theta^0 \,P_0 e^{r^{c,D}\, \tau } )\mathbbm{1}_{\{\tau<\tau^B\}} \right] = P_0(1-\theta^0)\,. 
\end{equation}
\end{Theorem}
\begin{proof} 
  Using optional stopping theorem and the fact that   $P e^{r^{c,E} -r^{c,D}}$ is $\mathbb Q$-martingale, for any $\tau \in \mathcal T$,  
\begin{equation}\label{eq loans overpriced}
\begin{split}
& \mathbb E^{\mathbb Q} \left[ e^{-r^{c,D} \tau}(P_\tau \,e^{r^{c,E}\,\tau } -\theta^0 \,P_0 e^{r^{c,D}\, \tau } )\mathbbm{1}_{\{\tau<\tau^B\}} \right] \\
& \leq 
\mathbb E^{\mathbb Q} \left[ e^{-r^{c,D} {\tau \wedge \tau^B}}(P_{\tau \wedge \tau^B} \,e^{r^{c,E}\,{\tau \wedge \tau^B} } -\theta^0 \,P_0 e^{r^{c,D}\, {\tau \wedge \tau^B} } ) \right]
= P_0(1-\theta^0)\,. 
\end{split}
\end{equation}
That means that the non-arbitrage price of the lending contract is always less or equal to the initial premium $P_0(1-\theta^0)$ and hence it is optimal to exercise at time $\tau^*=0$, i.e. to never enter the contract in the first place. 
\end{proof}

\begin{figure}
    \centering
    \includegraphics[width=0.5\linewidth]{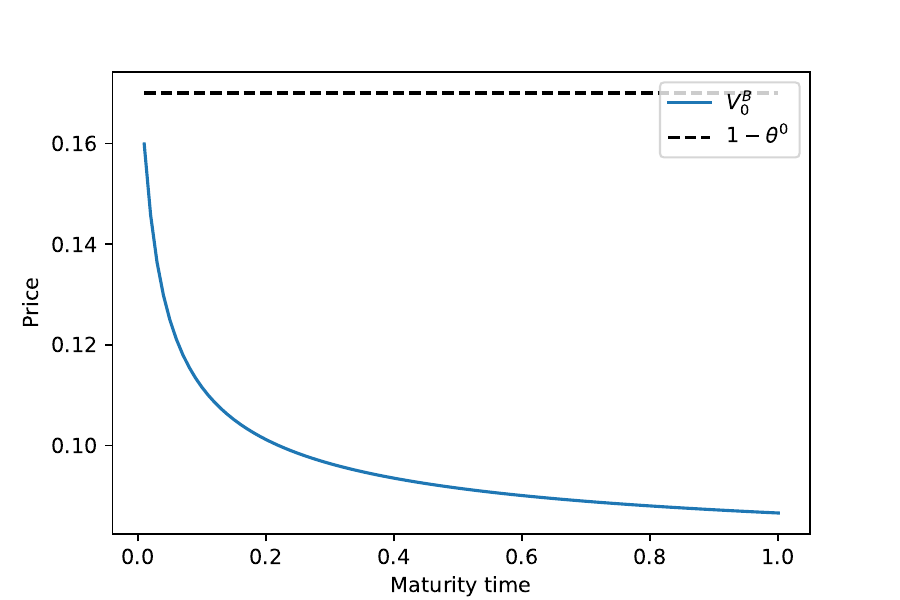}
    \caption{Comparison of European Barrier option versus $(1-\theta^0)$ for different values of $T\in(0,1], P_0=1, \sigma=0.5, \theta^0=0.83, \theta=0.9$.}
    \label{fig comparison}
\end{figure}

From the inequality \ref{eq loans overpriced} we see that market participants that enter lending contract position pay premium for the the additional optionality of exiting the contract at any time of their choosing. If that optionality was removed the lending contract would correspond to a European barrier option. For the purpose of demonstration in Appendix \ref{sec price formula without spread} we find a closed form for the price of this European barrier option when $P$ is modelled by geometric Brownian motion. In Figure \ref{fig comparison} we show how much lower is the value of the European barrier option in comparison to the American option price $P_0(1-\theta^0)$.

Next, we study the impact of liquidations and the time of exiting the loan contract on its non-arbitrage value under the Assumption \ref{as no spread}.

\subsection*{If there were no liquidations, keep the loan forever}
If there is no liquidation constraint, i.e., $\tau^B=\infty$,  the non-arbitrage price is given by  

\begin{align}
\sup_{\tau \in \mathcal T} \mathbb E^{\mathbb Q} \left[ (P_\tau \,e^{(r^{c,E}-r^{c,D})\,\tau } -\theta^0 \,P_0   )^{+} \right]\,,     
\end{align}
As $p\mapsto \varphi(p)\coloneqq (p-\theta^0 P_0)_+$ is convex,   by Jensen's inequality for conditional expectations,
for any stopping times $\tau, \rho \in \mathcal T$, s.t $\rho \geq \tau$
\begin{equation}
\begin{split}
    & \mathbb{E}[\varphi(P_\tau \,e^{(r^{c,E}-r^{c,D})\,\tau })]
    =\mathbb{E}[\varphi(\mathbb{E}[P_\rho \,e^{(r^{c,E}-r^{c,D})\,\rho }\mid \mathcal F_\tau])] \\
   & \le 
     \mathbb{E}[\mathbb{E}[\varphi(P_\rho \,e^{(r^{c,E}-r^{c,D})\,\rho })\mid \mathcal F_\tau]]=\mathbb{E}[\varphi(P_\rho \,e^{(r^{c,E}-r^{c,D})\,\rho })]\,,
     \end{split}
\end{equation}
where in the first equality we use the martingale property of $(e^{-r^{c,D}} P_t e^{r^{c,E} t})_t$. This shows that the value of such a contract is increasing with time. In particular taking $\tau=0$ we have that 
\begin{equation}
 \mathbb{E}[\varphi(P_\rho \,e^{(r^{c,E}-r^{c,D})\,\rho })]
    \geq  P_0(1-\theta_0)\,.
\end{equation}



\section{Borrowing and lending with rate spreads} 

In this section, we analyse pricing and hedging  \emph{from the borrower's perspective} of a loan contract under different rates for borrowing and lending. Classical non-arbitrage pricing presented in section \ref{no spread} no longer applies in this setting, and non-linear pricing theory, see \cite{el1997backward,dumitrescu2019optional} is required. 
The fair price of the loan contract is defined as the minimal endowment to finance a super-hedging strategy for the payoff.
In this section we work under the following assumption: 
\begin{Assumption} \label{as spread}
 We assume that there exists an external market where:
\begin{itemize}
\item The agent can borrow and lend any amount of USDC at the riskless rates $r^{b,D}$ and $r^{c,D}$, respectively.
\item The agent can buy and sell any amount of risky asset with price $P$. The risky asset earns interest/yields/dividends at the rate  $r^{c,E}$. To short, agents borrow the asset at the external market at the rate $r^{b,E}$\footnote{We make this assumption to prevent arbitrage that would result from borrowing risky assets with no interests while accruing interest on the lending platform. We are implicitly assuming the borrowing is taking place at the external market. The case of borrowing from the lending platform for the purpose of building (super) replicating portfolio would result in opening another lending contract, rendering the problem highly non-linear }.
\item The above transactions do not incur any transaction costs and the size of a trade does not impact the prices of the traded assets. 
\end{itemize}
\end{Assumption}

 We start by deriving the wealth dynamics.  
Let  $(V_t)_{t \in [0,T]}$ be the wealth process   (in USDC) and
$(\pi_t)_{t \in [0,T]}$ 
be the   process
representing the number of units 
 invested in ETH. Let  $\tau \in \mathcal T$ be a stopping time 
at which  the holder chooses to pay back the loan. The agent's  payoff at 
$\tau$ is then given by 
\begin{equation}
\label{eq:payoff_tau}
 \psi(\tau, P_\tau) := (P_\tau \,e^{r^{c,E}\,\tau } -\theta^0 \,P_0 e^{r^{b,D}\, \tau } )\mathbbm{1}_{\{\tau<\tau^B\}}.
\end{equation}
The aim of this section is to develop a deep neural network-based algorithm for learning trading strategies on the external markets that allow us to replicate the payoff  of the lending contracts that are not necessarily optimally exercised and assuming trades can only happen on a discrete time grid\footnote{this assumption is consistent with a DeFi protocol where discrete times correspond to block mining times.}. 
\begin{Definition}[Replicating portfolio]
If there exists the initial wealth process $V_0$ and strategy
$(\pi_t)_{t \in [0,T]}$ 
representing the number of units 
 invested in ETH such that  
\begin{equation} \label{replicating condition}
    V_{t} =  \psi(t, P_{t})\,, \quad \text{for all} \quad t \in \Pi \coloneqq \{ k \Delta t \mid k\in \mathbb N   \}\,,
\end{equation}
then we call $(V_t)_{t \in [0,T]}$  a replicating portfolio. 
\end{Definition}
Condition \eqref{replicating condition} ensures no matter when borrower decides to close the lending contract value of $V$ is sufficient to cover the liability\footnote{This is a conservative approach that does not require solving optimal stopping problem for the borrower. }.

Recall the notation $x^+ = \max\{0,x\}$ and $x^- = - \min\{0,x\}$. For any $t<\tau$,
given $V_{t}$ and $\pi_t$,
the value of wealth at ${t+\Delta t }$ for a small $\Delta t$ is given by
\begin{equation}\label{eq:wealth dynamics discrete}
\begin{split}
     V_{t+\Delta t} &= (1 + r^{c,D} \Delta t)(V_t - \pi_tP_t)^+ - (1 + r^{b,D} \Delta t)(V_t - \pi_t P_t)^- \\
    &\qquad + (1 + r^{c,E} \Delta t) \left(\pi_t \right)^+ (P_t + \Delta P_t) - (1 + r^{b,E} \Delta t) \left(\pi_t \right)^- (P_t + \Delta P_t)\,.
\end{split}    
\end{equation}
The first term is  due to  the interest rate earned by providing/holding USDC collateral, 
the second term is  due to the interest rate paid for  borrowing USDC collateral, 
the third term is   the value of wealth  at $t+\Delta t$ due to holding   ETH,
and the last term is the cost due to shortselling borrowed ETH.

As $x=x^+-x^-$,
\begin{equation}\label{eq replicating portfolio}
    \begin{split}
    V_{t+\Delta t} &= (1 + r^{c,D} \Delta t)(V_t - \pi_t P_t)  - (  r^{b,D}-r^{c,D}) \Delta t (V_t - \pi_t P_t)^-    \\
    &\quad + (1 + r^{c,E} \Delta t)  \pi_t   (P_t + \Delta P_t) - (  r^{b,E} -r^{c,E})\Delta t \left(\pi_t \right)^- (P_t + \Delta P_t)
    \\
  &=  (V_t - \pi_t P_t)   + r^{c,D} \Delta t  (V_t - \pi_t P_t)  - (  r^{b,D}-r^{c,D}) \Delta t (V_t - \pi_t P_t)^- \\
    &\quad + 
 {\pi_t}   \left(P_t + \Delta P_t\right)  + r^{c,E} \Delta t   {\pi_t}   \left(P_t + \Delta P_t\right) - (  r^{b,E} -r^{c,E})\Delta t \left( {\pi_t}  \right)^- \left(P_t + \Delta P_t \right)
 \\
  &=   V_t   + {\pi_t}    \Delta P_t   +
  r^{c,E} \Delta t   {\pi_t}     \Delta P_t
  - (  r^{b,E} -r^{c,E})\Delta t \left( {\pi_t}  \right)^-  \Delta P_t  
  \\
    &\quad + 
    \Delta t \left(r^{c,D}  (V_t - \pi_tP_t)  - (  r^{b,D}-r^{c,D})   (V_t - \pi_tP_t)^- 
  + r^{c,E} \pi_t P_t - (  r^{b,E} -r^{c,E}) \left( \pi_t \right)^- P_t  
  \right)\,.
    \end{split}
\end{equation}
We model the price of ETH by a discrete geometric Brownian Motion
\[
 P_{t_+\Delta t} =  P_{t}\,\exp\left( (\mu_{t} +\frac1{2}\sigma_t^2) \Delta t + \sigma_{t}\, \Delta W_{t}\right), 
    \]
where $\mu$ and $\sigma$ are bounded measurable functions such that $\inf_{t\in [0,T] }\sigma_t >0$. Plugging that into \eqref{eq replicating portfolio} we have  
\begin{equation}\label{eq replicating portfolio with P}
    \begin{split}
    &   V_{t+\Delta t} - V_{t}  \\
  &  =
      \left(r^{c,D}  V_{t}    - (  r^{b,D}-r^{c,D})   (V_{t} - \pi_{t} P_{t})^- 
  + (r^{c,E} -r^{c,D} +\mu_{t} ) \pi_{t} P_{t}  - (  r^{b,E} -r^{c,E}) \left( \pi_{t}   \right)^- P_{t}  
  \right)\Delta t
  \\
  &\quad 
  + \pi_t P_{t} \sigma_t \Delta W_{t} + r^{c,E} \Delta t   {\pi_{t}}     \Delta P_{t},
    \end{split}
\end{equation}

We see that from \eqref{eq replicating portfolio with P} that when there is no spread the no-linear terms of this equation disappear.

In Appendix \ref{sec non linear bsde pricing} we derive the nonlinear FBSDE from \eqref{eq replicating portfolio with P} for the fair price of the lending contract with non-zero rate spread \emph{from the borrower's perspective}. Explicit solution of the derived equation doesn't exist and one would need to resort its numerical approximation. Since our problem is path-dependent (due to the barrier) the equation is not Markovian making numerical simuations highly not-trivial and computationally intense. Instead in the next section we propose efficient deep learning algorithm by taking conservative approach and not optimising over the stopping times.

\subsection{Deep hedging.}

In this section we seek a pair of $(V_0,\pi)$ that minimises the hedging error $V_t - \psi_t$ by extending  deep hedging framework \cite{buehler2019deep} to the case of perpetual contracts. The framework allows one to incorporate transactional costs. Define the total cost of the trading strategy $\pi$ as
\[
C_{\tau}(\pi) := \sum_{t\in \Pi^n, t\leq \tau} c(\pi_{t} - \pi_{t-\Delta t}),
\]
where $c : \mathbb R \rightarrow \mathbb R_+$ is a constant fee paid for portfolio rebalancing. 

Next, we parametrise $\pi_t$ by a recurrent neural network such that $\pi_{t+\Delta t} \approx \pi^{\phi^*}(P_{t+\Delta t}, \pi_{t})$ with $\phi^* \in \mathbb R^p, v_0^*\in \mathbb R$ the network's parameters and the initial wealth value satisfying
\begin{equation}\label{eq opt deep hedge}
(\phi^*, v_0^*) := \arg \inf_{\phi, v_0} \quad \sum_{t\in\Pi^n} \mathbb E\left[\psi(t, P_{t}) - (V_{t} - C_{t}(\pi^\phi)) \right]^2.
\end{equation}

The above optimisation obtains the initial wealth $v_0^*$ and the hedging strategy $\pi^{\phi^*}$ such that at every time step $t\in\Pi^n$ the wealth process $V_{t}$ hedges the payoff of the Barrier option, accounting for transaction costs, \textit{regardless of the exercise time of the option}.

 

\subsubsection*{Experiment 1 - fixed spread and fixed cost}
We take the following parameters and the following modelling assumptions:
\begin{itemize}
    \item[--] We take the rates to be $r^{b,D}=0.12, r^{c,D}=0.08, r^{b,E}=0.025, r^{c,E}=0.017$. These rates taken from \url{https://app.aave.com/markets/} on the 1st of April 2024.
    \item[--] We take horizon time $T=73$ days (one fifth of a year). 
    \item[--] We model the price process $P$ by a Geometric Brownian Motion
    \[
    P_{t+\Delta t} = P_t \, \exp\left(\mu - \frac1{2}\sigma^2)\Delta t + \sigma \Delta W_{t}\right), 
    \]
    and we repeat the optimization \eqref{eq opt deep hedge} for $\mu=-0.3,0.0,0.3$ (i.e. bearish, neutral and bullish market) and $\sigma=0.1,0.3,0.5$ (i.e. different volatility regimes).
    \item[--] We fix $\theta = 0.9$ and we take $\theta^0\in[0.8,0.9)$, that is we explore the price for different initial loan-to-values.  
    \item[--] We assume that a swap in Uniswap (necessary to buy or sell units of ETH) costs on average \$20, \cite{adams2023costs}.
\end{itemize}

Additionally, we compare the performance of $\pi^{\phi^*}$ against the performance of the delta-hedge trading strategy presented in section \eqref{sec delta hedge}.

\begin{figure}
    \centering
    \includegraphics[width=0.8\textwidth]{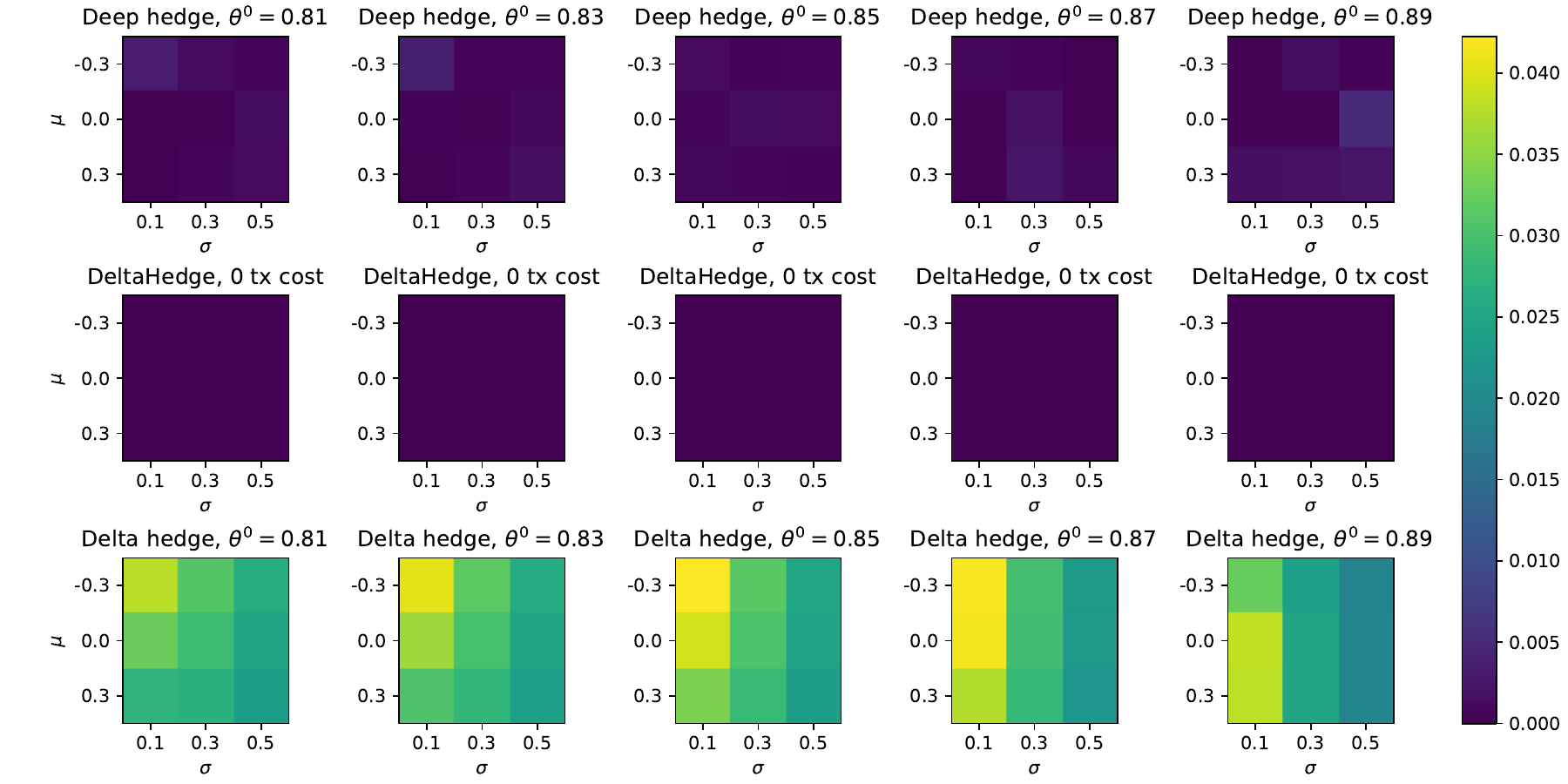}
    \caption{ Mean relative error $\frac1{n}\sum_{t\in\Pi^n} \mathbb E\left[\frac{\psi(t, P_{t}) - (V_{t} - C_{t}(\pi))}{\psi(t, P_{t})} \right]$ for (a) $\pi=\pi^{\phi^*}$ and including transaction cost (first row), (b) $\pi=\Delta(t, P_t)$ and 0 transaction cost (second row) and (c) $\pi=\Delta(t, P_t)$ and including transaction cost.  Each column is a different value the initial loan to value $\theta^0$. Heatmaps'axis correspond to volatility and market regime. }
    \label{fig rel error}
\end{figure}

Figure \ref{fig rel error} provides the mean relative error $\frac1{n}\sum_{t\in\Pi^n} \mathbb E\left[\frac{\psi(t, P_{t}) - (V_{t} - C_{t}(\pi))}{\psi(t, P_{t})} \right]$ for $\pi=\pi^{\phi^*}$ (first row) and for $\pi=\Delta(t, P_t)$ for different values of initial loan to value $\theta^0$, volatility and market regime. The learned trading strategy has a relative error of the order $O(10^{-3})$ (less than 1\%), one order of magnitude smaller than the delta-hedge trading strategy when considering transaction costs. Additionally, Figure \ref{fig rel error} compares the relative error learned trading strategy against the delta-hedge when there are 0 transaction costs, yielding similar relative error than

\begin{figure}
    \centering
    \includegraphics[width=0.5 \textwidth]{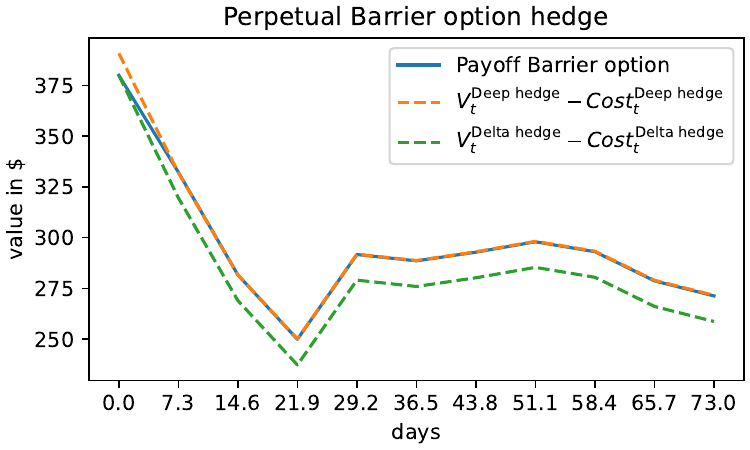}
    \caption{Payoff of the Barrier option, wealth process $V_{t}$ using $\pi^{\phi^*}, v_0^*$, and wealth process using the delta hedge and $P_0(1-\theta^0)$ for one random seed and combination of parameters $\mu=0., \sigma=0.1, \theta^0=0.83$}
    \label{fig hedge}
\end{figure}

In Figure \ref{fig hedge} we compare the payoff $\psi(t, P_t)$ with the wealth process $V_{t}$ using (a) $v_0^*$ and $\pi^{\phi^*}$ (b) $P_0(1-\theta^0)$ and the delta hedge. The hedging strategy $\pi^{\phi^*}$ is able replicate $\psi(t, P_t)$ as it is trained to account for transaction costs and the non-zero spread between supply and borrow interest rates.

\begin{figure}
    \centering
    \includegraphics[width=0.8 \textwidth]{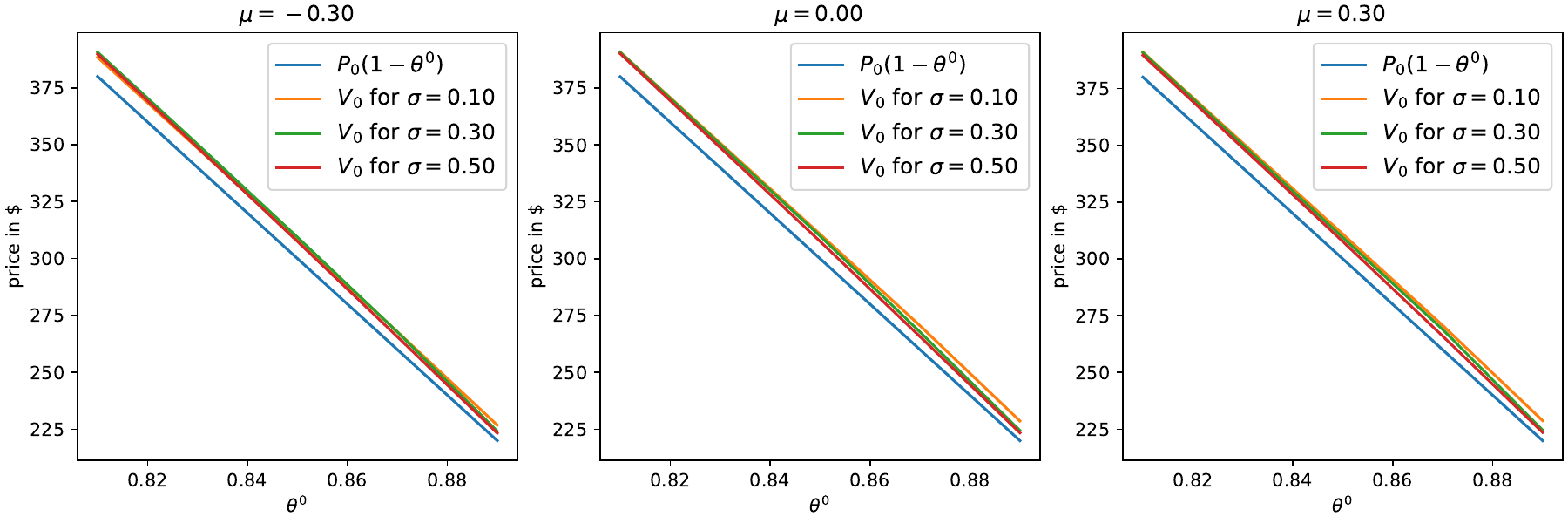}
    \caption{Comparison between initial price of the loan, $P_0(1-\theta^0)$, and initial price of the replicating portfolio $v_0^*$ for different market regimes. }
    \label{fig price}
\end{figure}

In Figure \ref{fig price} we compare the  $v_0^*$, initial wealth to enter the portfolio $\pi^{\phi^*}$, against the loan price $P_0(1-\theta^0)$ in the different market regimes. We see that, due to transaction costs, $v_0^*$ is above $P_0(1-\theta^0)$. The market regime does not seem to impact the value of $v_0^*$.

\begin{figure}
    \centering
    \includegraphics[width=0.5 \textwidth]{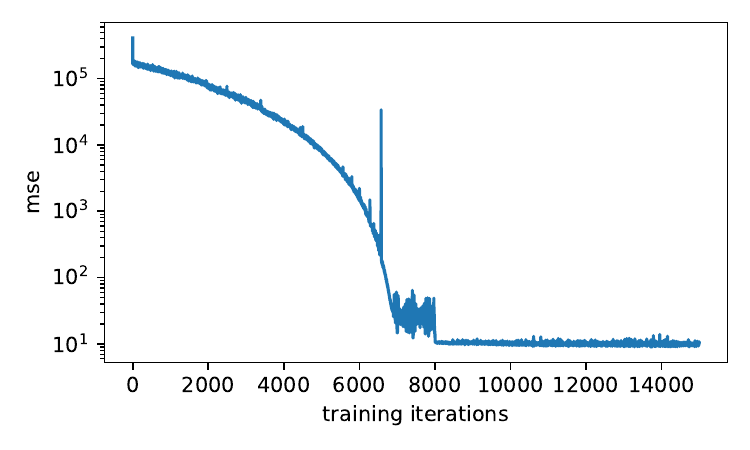}
    \caption{Training error of \eqref{eq opt deep hedge} for one combination of parameters $\mu=-0.3, \sigma=0.1, \theta^0=0.83$.}
    \label{fig training error}
\end{figure}

Finally, Figure \ref{fig training error} provides the training error of \eqref{eq opt deep hedge} for one combination of parameters $\mu=-0.3, \sigma=0.1, \theta^0=0.83$. 

\subsubsection*{Experiment 2 - Variable spread and variable transaction cost}
In this experiment we study the contribution of the spread and the transaction costs to the price and hte hedging relative error. We take the following parameters:
\begin{itemize}
    \item[--] We take the rates $r^{c,D} = 0.08, r^{b,D} = r^{c,D} + \text{spread}, r^{c,E} = 0.017, r^{b,E} = r^{c,E} + \text{spread}$ where $\text{spread} \in [0,0.5]$.
    \item[--] We take horizon time $T=73$ days (one fifth of a year).
    \item[--] We model the price process $P_{t}$ by a Geometric Brownian Motion
    \[
    P_{t + \Delta t} = P_t + \mu P_t \Delta t + \sigma P_{t} \Delta W_{t}
    \]
    and fix $\mu = 0, \sigma = 0.8$.
    \item[--] We fix $\theta = 0.9$ and $\theta^0=0.83$. 
    \item[--] We repeat the the optimization \eqref{eq opt deep hedge} for zero transaction costs, and for Uniswap swaps costing in average \$20. 
\end{itemize}

\begin{figure}
    \centering
    \includegraphics[width=\textwidth]{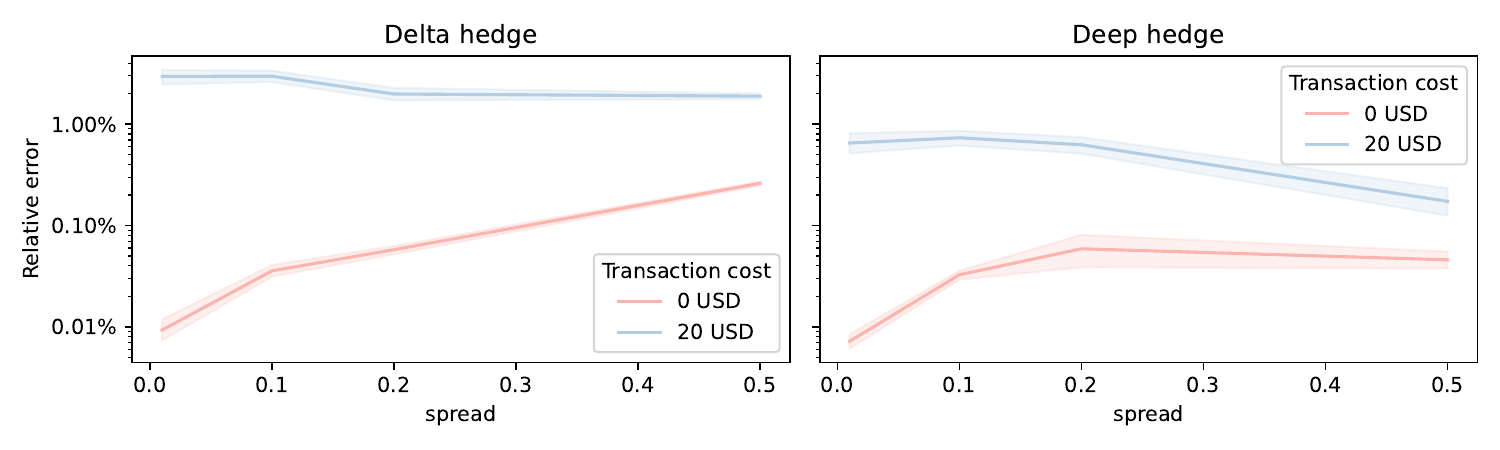}
    \caption{Relative error of delta hedge and deep hedge for fixed parameters $\mu=0, \sigma=0.8, \theta^0=0.83$, and variable transaction cost and rate spreads.}
    \label{fig cost spread}  
\end{figure}

Figure \ref{fig cost spread} shows the relative error of the hedging strategies in terms of the spread and the cost. The relative error $\frac1{n}\sum_{t\in\Pi^n} \mathbb E\left[\frac{\psi(t, P_{t}) - (V_{t} - C_{t}(\pi))}{\psi(t, P_{t})} \right]$ is calculated 10 times over simulated Monte Carlo samples of size $100\,000$. The deep hedge with transaction cost is one order of magnitude better than the delta hedge. In the case of the delta hedge, we see that the spread accounts for $O(1/100)$ of the relative error when the spread is around 0.1 which is the realistic scenario.

\begin{figure}
    \centering
    \includegraphics[width=0.5\textwidth]{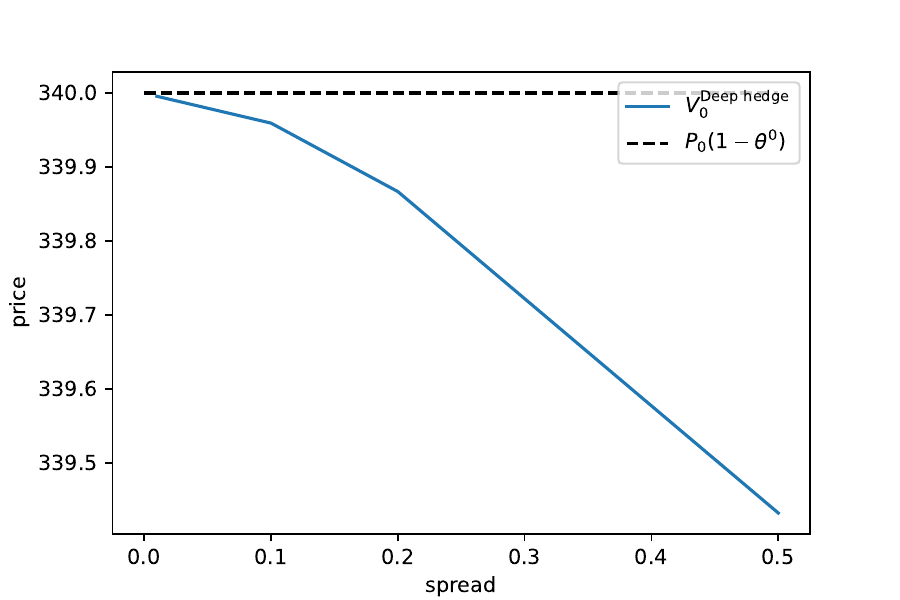}
    \caption{Price of the lending option derived from the deep hedge for fixed parameters $\mu=0, \sigma=0.8, \theta^0=0.83$, 0 transaction costs, and variable rate spread $\in[0,0.5]$.}
    \label{fig price spread}  
\end{figure}

Figure \ref{fig price spread} shows the price of the lending option derived from the deep hedge as a function of the rate spread. 

\subsection{Delta hedging.}\label{sec delta hedge}

Here we seek a simple and approximate solution for the pair $(V_0,\pi)$ 
for evaluation of lending contracts. We can derive the dynamics of $t\mapsto \psi(t,P_t)$, for $t <\tau^B$, using Taylor expansion,

\begin{equation}\label{eq perp}
    \begin{split}
        \Delta \psi_{t}  & = \psi(t+\Delta t, P_{t+\Delta t}) - \psi(t, P_{t}) \\
        & = \partial_t \psi(t, P_{t}) \Delta t + \partial_x \psi(t, P_{t}) \Delta P_t + \partial_{t,x} \psi(t, P_{t}) \Delta t \Delta P_t + O(\Delta t^2) \\
        & = \left(\partial_t \psi(t, P_{t}) + \partial_x \psi(t, P_{t}) P_t \mu_{t} \right) \Delta t + \partial_x \psi(t, P_{t}) \sigma_{t} P_{t} \Delta W_t + \partial_{t,x} \psi(t, P_{t}) \Delta t \Delta P_{t} + O(\Delta t^2).
    \end{split}
\end{equation}

By comparing the terms multiplying $\Delta W_t$ and $\Delta t\Delta P_t$ in \eqref{eq perp} and \eqref{eq replicating portfolio with P} we set
\begin{equation}
    \Delta(t,P_t) = \partial_x \psi(t, P_t) =  e^{r^{c,E}t}\,, \quad \text{for} \quad   t <\tau^B\,.
\end{equation}
Furthermore, in the following we set $V_0 = \psi(0,P_0) = P_0(1-\theta^0)$. With this choice of the pair $(V_0, \pi)$, we can calculate the difference of the increments $\Delta V_t - \Delta \psi_t$. 

Note that since $V_0 = \psi_0$, if their increments were the same then $V_t$ and $\psi_t$ would also be the same for all $t$. This would imply that the initial amount of cash $V_0$ together with the trading strategy $\pi$ would allow to build a portfolio replicating the payoff of the lending contract for all $t < \tau^B$. 
\[
\begin{split}
&\Delta V_t - \Delta \psi_t \\
& =  \left(r^{c,D}  V_t    - (  r^{b,D}-r^{c,D})   (V_t - \pi_t P_t)^- 
  + (r^{c,E} -r^{c,D} +\mu_t ) \pi_t P_t  - (  r^{b,E} -r^{c,E}) \left( \pi_t   \right)^- P_t  
  \right)\Delta t \\
  & \quad - \left(\partial_t \psi(t, P_t) + \partial_x \psi(t, P_t) P_t \mu_t \right) \Delta t,
\end{split}
\]
which is not 0 almost surely. Hence the pair $V_0 = P_0(1-\theta^0))$ and  $\Delta(t,P_t)$ is not a   replicating portfolio, but as we see from simulations provides its approximation.  



We see that in general, when $V_0=P_0(1-\theta^0)$ there is no $\pi$ such that replicating condition holds unless $r^{c,\cdot} = r^{b, \cdot} = 0$.  Using simulations that we present in the section (Figure \ref{fig cost spread}) we study the error of $V_t - \psi_t$ when using $\pi_t = \partial_x \psi(t, P_t)$ and $V_0 = P_0(1-\theta^0)$ under various market conditions. 

The fact that $V_t - \psi_t$ is not zero for all $t$ can be understood through the analysis of perpetual options which include an additional \textit{streaming fee} needed to eliminate arbitrage opportunities (see \cite{cohen2023inefficiency} for an example on how to derive the non-arbitrage streaming fee for liquidity provision in AMMs). This streaming fee is not included in the current design of lending protocols.

\subsection{Covered barrier - Protocol perspective.}

When an agent opens long-ETH lending position, the balance sheet of a lending protocol gains $1$ ETH and loses $\theta^0 P_0$ of USDC. 
Note that $ - \theta^0 P_0 = P_0(1-\theta^0) - P_0 $ where $P_0(1-\theta^0)$ corresponds to option premium and $- P_0$ cost of buying $1$ ETH (static hedge). The  $1$ ETH is kept in the pool and hence when there is rehypothecation of the collateral it earns interests. Before the liquidation, that is when $t<\tau^B$, lending protocol holds a \textit{covered call option} on it's balance sheet that has the payoff
\begin{equation}
    -(P_t\,e^{r^{c,E}\,t} - \theta^0 \,P_0\,e^{r^{b,D}\, t}) + \,e^{r^{c,E}} P_t 
    =  \theta^0 \,P_0\,e^{r^{b,D}\,t} \,.
\end{equation}
We see that prior to liquidation, the protocol is delta-neutral with respect to the change in the price of ETH.  However, when the position becomes open for a liquidation which happens when for the first time $ P_t \leq \theta ^{-1}\theta^0 \,P_0\ e^{(r^{b,D} -r^{c,E})  \,t}$ multiple outcomes are possible. The position can be fully liquidated, which essentially is equivalent to the loan being fully paid off. The liquidation is not successful, leading to the payoff for the protocol being $e^{r^{c,E}} P_t$ (and hence exposure to a risky asset). Only part of the position is liquidated, leading to some amount of borrowed and collateral assets staying on the protocol balance sheets.

\bibliographystyle{siam}
\bibliography{references.bib}

\appendix

\section{Analytic formula without borrowing and lending rate spreads}\label{sec price formula without spread} 

  This section presents the analytic formula for  the buyer's price in the case without interests rate spreads, i.e.,  
   $r^{b,D}=r^{c,D}$ and 
  $r^{b,E}=r^{c,E}$, considering the case where the buyer takes the stopping time $\tau =T$ instead of  strategically optimising  over all stopping times $\tau$.

First we observe that interest paid on collateral, represented here by risky asset $P$, are equivalent to the asset paying a dividend with the rate $r^{c,E}$. Classical non-arbitrage pricing theory applied to risky asset yielding dividends, tells us that that absence of arbitrage implies the existence of martingale measure $\mathbb Q$ such that $(e^{-r^{c,D}t} P_t e^{r^{c,E} t})_t$ is $\mathbb Q$-martingale. We will next find such measure $\mathbb Q$.

Let $W=(W_t)_{t\ge 0}$ be a standard Brownian motion on a probability space 
$(\Omega, \mathcal F,\mathbb{P})$
and let $\mathbb F=(\mathcal{F})_{t\ge 0}$
be the natural filtration of $W$ augmented with $\mathbb{P}$-null sets. 
 Fix
 $P_0>0$, 
 $\theta^0\in [0,1)$
 and 
 $ \theta\in (\theta_0,1)$. 
The price $P$ of ETH/USDC is given by 
\begin{equation}\label{eq gbm}  
    P_t=P_0 \exp\left( (\mu - \frac{1}{2}\sigma^2)t + \sigma W_t  \right)\,,\quad t>0\,,
\end{equation}
 for some $\mu\in \mathbb R$ and $\sigma>0$. We denote $Y_t := P_t e^{r^{c,E}t}$ the value of the risky asset with the compounded dividends at time $t$. Then
 \begin{equation}\label{eq gbm dividends}  
    Y_t=P_0 \exp\left( (\mu + r^{c,E} - \frac{1}{2}\sigma^2)t + \sigma W_t  \right)\,,\quad t>0\,,
\end{equation}
 Define the $\mathbb F$-martingale $(\gamma_t)_{t\geq 0}$ by
 \[
 \gamma_t = \exp\left( -\nu W_t - \frac1{2} \nu^2 t \right),\,\, \nu = \frac{\mu + r^{c,E} - r^{c,D}}{\sigma},\,\,t\geq 0 
 \]
and the measure $\mathbb Q$ with the Radon-Nikodym derivative
\begin{equation}\label{eq change of measure}
\frac{d \mathbb Q}{d \mathbb P}\bigg |_{\mathcal F_t} = \gamma_t, \,\, t\geq 0. 
\end{equation}
By the Girsanov theorem, the process $X = \{X_t := \nu t + W_t, t\geq 0\}$ is an $\mathbb F$-Brownian motion with respect to $\mathbb Q$. Note that under $\mathbb Q$,  $(e^{-r^{c,D}t} Y_t)_{t\geq 0}$ is a martingale,
\[
e^{-r^{c,D}t} Y_t = Y_0 \exp\left(-\frac1{2} \sigma^2 t + \sigma \underbrace{\left( \frac{\mu + r^{c,E} - r^{c,D}}{\sigma} t + W_t \right)}_{\text{Brownian motion under } \mathbb Q}  \right).
\]
The fair price of the option is
\begin{equation} 
\label{eq:value_P0 european}
\begin{split}
V(P_0) & :=  \mathbb{E}^\mathbb Q\left[ e^{-r^{c,D}T} (P_ \tau \,e^{r^{c,E}\, T} - \theta^0 \,P_0\,e^{r^{b,D}\, T} )\mathbbm{1}_{\{T<\tau^B\}}\right]\, \\
      & = \mathbb{E}^\mathbb Q\left[ e^{-r^{c,D}\,T} (Y_T  - \theta^0 \,P_0\,e^{r^{b,D}\, \tau} )\mathbbm{1}_{\{T<\tau^B\}}\right]\,,
\end{split}
\end{equation}
where 
\begin{equation*} 
\begin{split}
   \tau^B & :=   \inf  \{ t \geq 0: \theta\, \overbrace{P_t e^{r^{c,E}\, t}}^{Y_t} = \theta^0 {P_0} e^{r^{b,D}\, t}   \} \, \\
          & =  \inf  \left\{ t \geq 0: \frac{Y_t}{P_0} e^{-  r^{b,D} \, t} = \theta^{-1 }\theta^0  \right\} \,
\end{split}
\end{equation*}

 Observe that by setting $S_t :=\frac{Y_t}{P_0} e^{-  r^{b,D}\, t}$ and considering no-spread $r^{c,D} = r^{b,D}$,
\eqref{eq:value_P0 european}  can be equivalently written as 
 \begin{equation} 
 \label{eq:value_S0 european}
V(P_0) := P_0 \mathbb{E}^\mathbb Q\left[(S_T - \theta^0   )\mathbbm{1}_{\{T<\tau^B\}}\right]\,,
\end{equation}
where
\begin{equation} 
\label{eq:S_dynamics}
S_t =  \exp\left(-\frac1{2} \sigma^2 t + \sigma \underbrace{\left( \frac{\mu + r^{c,E} - r^{c,D}}{\sigma} t + W_t \right)}_{\text{Brownian motion under } \mathbb Q}  \right),
\end{equation}
and 
\begin{equation}
\label{eq:tau_B_S}
   \tau^B :=   \inf  \left\{ t \geq 0: S_t = B    \right\} \,,
\end{equation}
where $B\coloneqq \theta^{-1 }\theta^0 \in (\theta^0,1)$.

The price \eqref{eq:value_S0 european} is an European  down-and-out call option where
the underlying asset starts at $S_0=1$ and pays no dividends,
the interest rate is zero, and the barrier is  
a barrier
larger than the strike. One can show by
  the reflection principle that 
\begin{align}
\label{eq:do_european_call}
\begin{split}
&\mathbb{E}^\mathbb Q[(S_T    -\theta^0  )_{+}\mathbbm{1}_{\{T<\tau^B\}}]
\\
&= C_v(1, T,B)+(B-\theta^0)C_d(1,T, B)-\frac{1}{B}\left(C_v(B^2, T,B)+(B-\theta^0)C_d(B^2,T, B) \right),
\end{split}
\end{align}
where $C_v$ is the vanilla call price given by
\begin{align*}
C_v (S,T,E)=S N(d_1)-KN(d_2), \quad d_1=\frac{\log(S/E)+\frac{1}{2}\sigma^2T}{\sigma\sqrt{T}},\quad 
d_2=d_1-\sigma \sqrt{T},
\end{align*}
and $C_d$ is the digital call price given by
$$
C_d (S,T,E)= N(d_2).
$$
Substituting the formula of $C_v$ and $C_d$ into \eqref{eq:do_european_call}  yields that 
\begin{equation}\label{eq european barrier option price}
\begin{split}
&\mathbb{E}[(S_T    -\theta^0  )_{+}\mathbbm{1}_{\{T<\tau^B\}}]
\\
&=  N(\bar{d}_1)-B N(\bar{d}_2) +(B-\theta^0)N(\bar{d}_2) 
-\frac{1}{B}\left(B^2 N(\hat{d}_1)-BN(\hat{d}_2)  +(B-\theta^0)N(\hat{d}_2)   \right)
\\
&=  N(\bar{d}_1)-\theta^0 N(\bar{d}_2) 
- B N(\hat{d}_1)+ \frac{\theta^0}{B}  N(\hat{d}_2),
\end{split}
\end{equation}
where 
\begin{align*}
\bar{d}_1 &= \frac{-\log(\frac{\theta^0}{\theta})+\frac{1}{2}\sigma^2 T}{\sigma\sqrt{T}}, \quad \bar{d}_2= \frac{-\log(\frac{\theta^0}{\theta})-\frac{1}{2}\sigma^2 T}{\sigma\sqrt{T}},
\\
\hat{d}_1 &= \frac{\log(\frac{\theta^0}{\theta})+\frac{1}{2}\sigma^2T}{\sigma\sqrt{T}}, \quad \hat{d}_2 =  \frac{\log(\frac{\theta^0}{\theta})-\frac{1}{2}\sigma^2T}{\sigma\sqrt{T}}.
\end{align*}
As  $\hat{d}_1=-\bar{d}_2$, $\hat{d}_2= -\bar{d}_1$ and $\theta^0/B=\theta$, 
\begin{align}
\begin{split}
&\mathbb{E}[(S_T    -\theta^0  )_{+}\mathbbm{1}_{\{T<\tau^B\}}]
\\
&=  N(\bar{d}_1)-\theta^0 N(\bar{d}_2) 
-B N(\hat{d}_1)+ {\theta}  N(\hat{d}_2)
\\
&=N(\bar{d}_1)-\theta^0 N(\bar{d}_2) 
-\frac{\theta^0}{\theta} (1- N(\bar{d}_2))+\theta (1-N(\bar{d}_1))
\\
&=\theta- \frac{\theta^0}{\theta} + (1-\theta) N(\bar{d}_1)+\theta^0(\frac{1}{\theta}-1) N(\bar{d}_2).
\end{split}
\end{align}
Consequently, 
with $\tau=T$, 
the buyer's price at $t=0$ is 
\begin{align}
V^B_0=P_0 \left(\theta- \frac{\theta^0}{\theta} + (1-\theta) N(\bar{d}_1)+\theta^0(\frac{1}{\theta}-1) N(\bar{d}_2)\right),
\end{align}
where 
\begin{align*}
\bar{d}_1 &= \frac{-\log(\frac{\theta^0}{\theta})+\frac{1}{2}\sigma^2 T}{\sigma\sqrt{T}}, \quad \bar{d}_2= 
\bar{d}_1-\sigma\sqrt{T}.
\end{align*}

Note that one expects that 
the European barrier option's price  \eqref{eq european barrier option price} is strictly less than the American option price $1-\theta^0$. 
Indeed, 
as discussed above,  for the American barrier option, it is optimal to stop before $S$ touches the barrier. This implies the stopping time $\tau= T$ for European barrier option  is strictly sub-optimal, as  there is non-zero probability for $S$ to touch  the barrier before $T$. Consequently, the European barrier option's price should be strictly smaller than the American one.

\section{Nonlinear pricing framework for the lending contract}\label{sec non linear bsde pricing}

In this section, we derive a   nonlinear backward equation for the fair price of the lending contract with non-zero rate spread \emph{from the borrower's perspective}.

As before the price process $P$ of ETH/USDC satisfies \ref{eq gbm}.

Furthermore, we consider the replicating portfolio of the lending contract payoff \eqref{eq:payoff_tau} with wealth process $V$ whose dynamics obey \eqref{eq replicating portfolio with P}. Sending $\Delta t \rightarrow 0$ yields 

\begin{equation*}
    \begin{split}
  \d V_t &  =
      \left(r^{c,D}  V_{t}    - (  r^{b,D}-r^{c,D})   (V_{t} - \pi_{t} P_{t})^- 
  + (r^{c,E} -r^{c,D} +\mu_{t} ) \pi_{t} P_{t}  - (  r^{b,E} -r^{c,E}) \left( \pi_{t}   \right)^- P_{t}  
  \right)\d t
  \\
  &\quad 
  + \pi_t P_{t} \sigma_t \d W_{t} ,
    \end{split}
\end{equation*}

Using a change of variable which
associates $\pi\,P \in \mathcal H^2(\mathbb R) $ with $Z= \pi\,P\,\sigma \in \mathcal H^2(\mathbb R) $, one can write the wealth dynamics as 
\begin{align}
\label{eq:wealth_dynamics}
   \d  V_{t }  
  &=   f(t,V_t, Z_t) 
     \d t
  + Z_t \d W_t,
\end{align}
where 
$$
f(t,y,z) =
 r^{c,D}  y    - (  r^{b,D}-r^{c,D})   (y - \frac{z}{\sigma_t} )^- 
  + \frac{ r^{c,E} -r^{c,D} +\mu_t }{\sigma_t} z    - 
  \frac{r^{b,E} -r^{c,E}}{\sigma_t}
   \left( z  \right)^-  .
$$
Thus the value of the lending contract from the buyer's perspective is given by 
$$
V^B_0=\sup_{\tau \in \cT} V^{P_0,\tau}_0\,,
$$
where 
 $V^{P_0,\tau}$
satisfies 
\eqref{eq:wealth_dynamics} for all $t\in   [0,\tau] $
and 
$V^{P_0,\tau}_\tau$ is given by 
\eqref{eq:payoff_tau}.

We now apply a Girsanov’s transform to remove the drift $\mu$. Indeed, we define the probability measure $\sQ$ equivalent to $\sP$ with Radon-Nikodym derivative \eqref{eq change of measure}. 
Then $W^\sQ= W_t+\frac{ r^{c,E} -r^{c,D} +\mu_t }{\sigma_t}t$ is a standard Brownian motion on the filtered space 
$(\Omega, \cF,\sF,\sQ)$. 
Hence under the measure $\sQ$, the buyer's optimal value is given by
$$
V^B_0=\sup_{\tau \in \cT} V^{P_0,\tau}_0\,,
$$
where 
$(P, V^{P_0,\tau}, Z^{P_0,\tau})$
is the unique square-integrable solution to the following forward-backward stochastic differential equation (FBSDE): for all $t\in [0,\tau]$,
\begin{align}
\label{eq:fbsde_p0}
\begin{split}
\d P_t &=  P_t \left( (r^{c,D}-r^{c,E})  \d t + \sigma_t \d W^\sQ_t \right),  
\\ 
\d  V_{t }  
  &=  - g(t,V_t, Z_t) 
     \d t
  + Z_t \d W^\sQ_t,
  \\
  V_\tau &= (P_\tau \,e^{r^{c,E}\,\tau } -\theta^0 \,P_0 e^{r^{b,D}\, \tau } )_{+}\mathbf{1}_{\{\tau<\tau^B\}},
\end{split}
\end{align}
 where $g $ is defined by 
 $$
 g(t,y, z)=
- r^{c,D}  y    + (  r^{b,D}-r^{c,D})   \left(y - \frac{z}{\sigma_t} \right)^- 
    +
  \frac{r^{b,E} -r^{c,E}}{\sigma_t}
   \left( z  \right)^-,
 $$
 and $\tau^B $ is defined as in   \eqref{eq:liquidation_time}.
 If $\tau^*\in \mathcal T$ is the optimal stopping time, then 
  the corresponding hedging strategy is given by
  $ \pi^*_t =Z^{P_0,\tau^*}_t/\left(P_t \sigma_t\right)$.

Finally, 
observe that $g$ is homogeneous in $y,z$ in the sense that for all $\alpha>0$,
$g(t,\alpha y,\alpha z)= \alpha g(t,y,z)$. Hence  
by introducing the scaled processes 
$$
(S_t , V^\tau_t, Z^\tau_t)= \left(\frac{1}{P_0} P_t e^{(r^{c,E}-r^{b,D})t},
 \frac{1}{P_0}V^{P_0,\tau}_t e^{-r^{b,D} t},
 \frac{1}{P_0}Z^{P_0,\tau}_t e^{-r^{b,D} t}
 \right),
 \quad t>0,
$$
 the buyer's price can be equivalently expressed as   
\begin{align}
\label{eq:v^B_P_0}
V^B_0=P_0\sup_{\tau \in \cT} V^{\tau}_0\,,    
\end{align}
where $(V^{\tau}, Z^{\tau})$ satisfies 
\begin{align}
\label{eq:fbsde}
\begin{split}
\d S_t &=  S_t \left( (r^{c,D}-r^{b,D})  \d t + \sigma_t \d W^\sQ_t \right), 
\quad 
S_0=1,  
\\ 
\d  V_{t }  
  &=  - \bar{g}(t,V_t, Z_t) 
     \d t
  + Z_t \d W^\sQ_t,
\quad 
  V_\tau  =  
 (S_\tau    -\theta^0  )_{+}\mathbf{1}_{\{\tau<\tau^B\}},
\end{split}
\end{align}
and $\bar{g}$ and $\tau^B$ are given by
\begin{align}
\label{eq:bar_g}
\bar{g}(t,y, z) &\coloneqq 
( r^{b,D} - r^{c,D})  y    + (  r^{b,D}-r^{c,D})   \left(y - \frac{z}{\sigma_t} \right)^- 
    +
  \frac{r^{b,E} -r^{c,E}}{\sigma_t}
   \left( z  \right)^-,
   \\
   \tau^B & \coloneqq \inf \left\{ t \in [0,T] \mid   S_t   \leq \theta^0/\theta     \right\}\,.
\end{align}




\end{document}